\documentclass[a4paper]{article}

\usepackage{lineno}
\usepackage{float}
\usepackage{multicol}
\usepackage{makecell}
\usepackage{multirow}
\usepackage{threeparttable}
\usepackage[margin=1in]{geometry} 
\usepackage{authblk}
\usepackage{hyperref}
\usepackage[T1]{fontenc}
\usepackage{amsmath}
\usepackage{amsthm}
\usepackage{amssymb}
\usepackage{lineno}
\usepackage{float}
\usepackage{multicol}
\usepackage{graphicx}
\usepackage{geometry}
\usepackage[utf8]{inputenc}
\usepackage[sort&compress,numbers,square]{natbib}
\usepackage{adjustbox}
\usepackage[font={footnotesize}]{caption}

\usepackage{abstract}

\title{Unveiling the dynamical diversity of quantum dot lasers subject to optoelectronic feedback}

\author[1]{Shihao Ding \thanks{These authors contributed equally to this paper}}
\author[1]{Shiyuan Zhao \thanks{These authors contributed equally to this paper}}
\author[2]{Justin Norman}
\author[2]{Bozhang Dong}
\author[1]{Heming Huang}
\author[2]{John Bowers}
\author[1,3]{Frédéric Grillot \thanks{frederic.grillot@telecom-paris.fr}}

\affil[1]{LTCI, T\'el\'ecom Paris, Institut Polytechnique de Paris, 19 Place Marguerite Perey, Palaiseau, 91120, France}
\affil[2]{Institute for Energy Efficiency, University of California, Santa Barbara, California 93106, USA}
\affil[3]{Center for High Technology Materials, University of New Mexico, 1313 Goddard St SE, Albuquerque, New Mexico 87106, USA}

\date{}
\begin{document}
\maketitle

\begin{abstract}
\noindent
This paper investigates experimentally and numerically the nonlinear dynamics of an epitaxial quantum dot laser on silicon subjected to optoelectronic feedback. Experimental results showcase a diverse range of dynamics, encompassing square wave patterns, quasi-chaotic states, and mixed waveforms exhibiting fast and slow oscillations. These measurements unequivocally demonstrate that quantum dot lasers on silicon readily and stably generate a more extensive repertoire of nonlinear dynamics compared to quantum well lasers. This pronounced sensitivity of quantum dot lasers to optoelectronic feedback represents a notable departure from their inherent insensitivity to optical feedback arising from reflections. Moreover, based on the Ikeda-like model, our simulations illustrate that the inherent characteristics of quantum dot lasers on silicon enable rapid and diverse dynamic transformations in response to optoelectronic feedback. The emergence of these exotic dynamics paves the way for further applications like integrated optical clocks, optical logic, and optical computing. 
\end{abstract}

\begin{multicols}{2} 
\section*{Introduction}
Silicon photonics is a swiftly evolving, important field to merge both photonics and electronics onto a silicon platform \cite{Atabaki2018, Shastri2021, Near2021}, which is a highly attractive solution for tackling the bottlenecks faced in high-volume datacom \cite{Near2021}. Notably, emerging applications using silicon photonic devices are also rapidly heating up, such as microwave generation \cite{chan2004tunable, juan2009microwave}, optical chaos cryptography \cite{abarbanel2001synchronization, gastaud2004electro}, optical sensing \cite{yao2016optoelectronic, wang2017optical}, and the optical computing \cite{larger2017high, Vandoorne2014, hou2018prediction, Shastri2021, Demkov21, Ashtiani2022, mourgias2023analog}. For instance, reservoir computing systems, operating with nonlinear dynamics generated by optical feedback and optical injection in semiconductor lasers, have been gaining significant traction and relevance over the past decade \cite{Vandoorne2014, bueno2017conditions, Gooskens22}. 

\begin{figure*}[t]
\centering{\includegraphics[width=150mm]{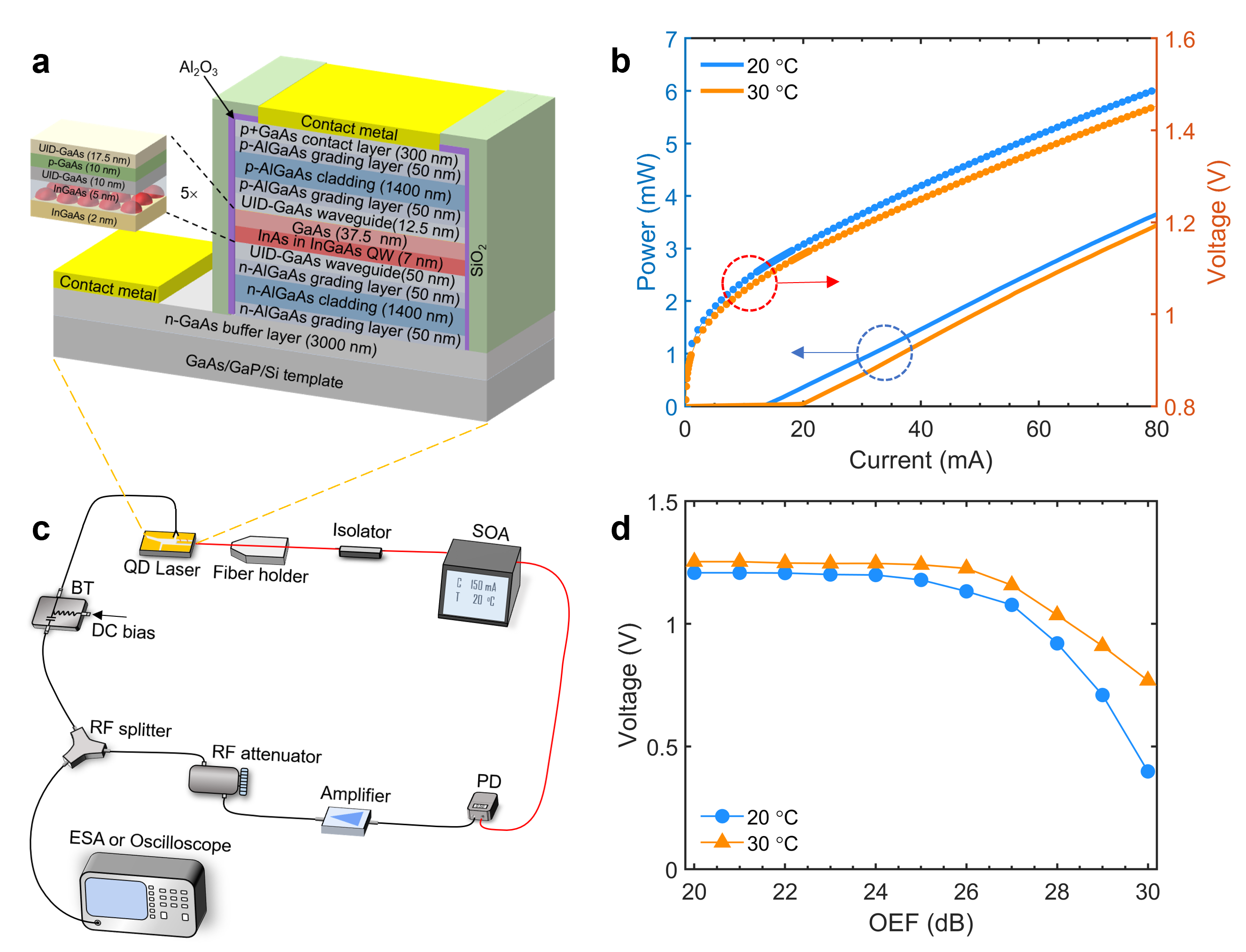}}
\caption{\textbf{Characterizations of the QD laser and the optoelectronic feedback setup.} \textbf{a} The epitaxial layer structure of the QD laser under study and \textbf{b} Optical power-voltage with bias current. \textbf{c} The tabletop experiment was used for studying optoelectronic feedback in the silicon-based QD laser (the solid red lines represent the optical path and the solid black lines represent the electrical path). \textbf{d} The measured voltage applied on the QD laser with respect to the OEF strength.}
\end{figure*} 

Recent works have demonstrated that nonlinear nodes in the reservoir layer can be emulated with the nonlinear dynamics of semiconductor lasers, especially in a delay-based photonic reservoir computing system \cite{Hulser22, Kanno22}. On this front, optoelectronic feedback (OEF) is an appealing method to activate nonlinear behaviors in laser diodes, as an OEF loop can be considered a closed-loop oscillator and is known for exhibiting diverse interesting dynamics. Such an oscillator, comprising a local optical path and a linear frequency-filtered electrical circuit, is capable of introducing a higher degree of behavioral complexity than that in traditional low-dimensional resonant systems. These conventional systems are governed by delay-differential equations (DDE) that effectively handle the complexity inside optoelectronic oscillation systems \cite{ikeda1979multiple, maleki2011optoelectronic}. In some research, the quantum well (QW) laser has been used to build an AC-coupled OEF loop with its inherent nonlinearity. However, this system has not yielded a diverse variety of dynamics, with predominantly periodic resonances or chaotic states being observed \cite{liu2002synchronized, islam2021optical}. In order to incorporate additional nonlinear elements into the OEF loop, the light emitted from the QW laser has been processed through a modulator, such as a Mach-Zehnder Modulator (MZM) and phase modulator (PM) \cite{bohm2019poor, chen2019reservoir, kouomou2005chaotic}. With such a configuration, diverse dynamics are introduced, including the emergence of previously unobserved mixed fast and slow states, often referred to as 'breathers'. This phenomenon is a direct result of the nonlinearity transfer function in the integrodifferential delay equation (iDDE) of the Ikeda-like model, manifesting in the sinusoidal response of the modulator \cite{kouomou2005chaotic,mbe2015mixed}. Although the inclusion of such additional nonlinear elements enhances the diversity of dynamics, it hampers the feasibility of silicon-based integration. This is primarily due to the increased complexity involved in integrating an electro-optical modulator. Considering factors such as cost, process simplicity, and stability control, it becomes imperative to identify an OEF loop that minimizes the number of elements while still preserving the potential for integration.

At this stage, quantum dot (QD) semiconductor lasers present themselves as enticing and promising candidates for laser sources compared to their QW counterparts. Leveraging the three-dimensional carrier confinement, QD lasers offer several key advantages that favorite their epitaxial integration onto silicon platforms, such as insensitivity to threading dislocation defects \cite{jung17tdd}, lower threshold current density \cite{bimberg97jstqe,lester99,deppe09}, and robust thermal stability \cite{maximov97,otsubo04}. In addition, QD lasers are renowned for their remarkable resilience against optical feedback, a characteristic that greatly benefits the design of isolator-free optical transmitters \cite{o2003feedback,chen2016electrically,huang2018analysis,arakawa22}. However, it is worth noting that their relatively high immunity to optical feedback might pose challenges when attempting to employ external control techniques for generating nonlinear dynamics. Therefore, the AC-coupled OEF loop incorporating a QD laser can be a highly versatile technique for exploring the intrinsic nonlinearity of the QD laser \cite{al2010excitability}. This approach can further drive diverse dynamics. Furthermore, this loop architecture exhibits high compatibility with the concurrent silicon-based processing of both the QD laser and the photon detector. Given the robustness and adaptability of the nonlinear dynamics it generates, this approach has considerable potential to catalyze breakthroughs in numerous fields, including but not limited to communication and sensing technologies.

In this paper, we investigate the nonlinear dynamic process of the optoelectronic feedback of an epitaxial QD laser on silicon at different temperatures. In addition, a QW laser (used as a reference laser) is employed in order to compare the dynamics under OEF with that from the epitaxial QD laser. Our analysis also incorporates the development of a complex numerical model based on the Ikeda model, which is used to explain the difference in dynamic generation by the nonlinearity transfer functions between QD and QW lasers. We believe that this work brings fresh information on QD laser physics operating under external control and can be used for various silicon photonic-based applications where optical computing, logic, clock, and microwave operations are required.

\section*{Results}
\subsection*{Description of the experimental setup}

\begin{figure*}[t]
\centering{\includegraphics[width=150mm]{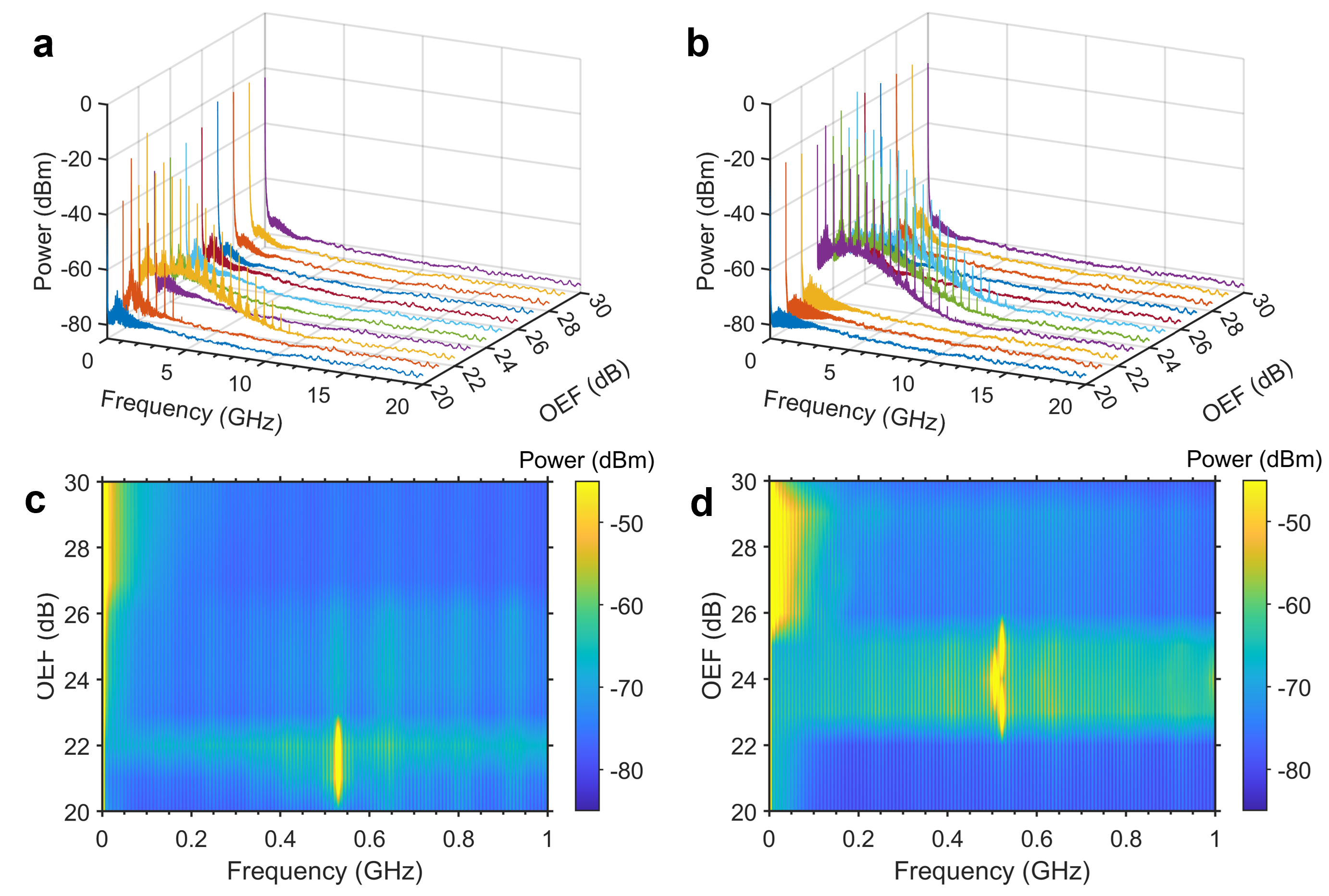}}
\caption{\textbf{Temperature impact on the dynamics of the QD laser with OEF.} Electrical spectra measured under different OEF strengths at \textbf{a} 20 {\textcelsius} and \textbf{b} 30 {\textcelsius} respectively. \textbf{c} and \textbf{d} are the corresponding mappings of the electrical spectra in the 1 GHz range at 20 {\textcelsius} and 30 {\textcelsius} respectively.}
\end{figure*} 

A silicon-based QD Fabry-Perot (FP) laser was used for the optical source in the OEF loop as shown in Figure 1(a). The optical output power-current-voltage (PIV) characteristics are shown in Figure 1(b) for temperatures of 20 {\textcelsius} and 30 {\textcelsius}, respectively. The threshold current is found to increase from 12 mA to 20 mA with temperature. Figure 1(c) displays the tabletop experiment used for studying OEF in the silicon-based QD laser. The feedback loop contains an optical isolator to eliminate back-reflections into the QD laser. While it may not be strictly necessary, its presence guarantees that the dynamics we present in this study arise from optoelectronic feedback rather than optical feedback. The output light from QD laser passes through a loop with optical (semiconductor optical amplifier) and electrical (photon detector, electrical amplifier, attenuator, power splitter, and bias tee) components. The selection of the amplifier aimed to investigate the properties of strong OEF by leveraging the cumulative saturation of the amplifier. On the other hand, the inclusion of the attenuator was strategically implemented to prevent any potential overloading of the laser circuit. The pump current ($J$) is set at twice the threshold for all measurements which is conveyed to the laser via the DC arm of the BT, while the feedback signal is transmitted through the AC arm. The electrical signal is then measured with an oscilloscope after the attenuator and before the BT. The low-noise amplifier within the PD possesses an inverting nature, and when combined with the amplifier within the electrical loop, it generates a cumulative negative feedback signal. The laser's architecture is anode grounded, which leads to an overall positive sign of the feedback since the injection-terminal voltage, as well as the feedback signal are both negative. The feedback strength is expressed in terms of the gain of the electrical amplifier in the OEF loop, with an overall delay $\tau$ of 250 ns. Figure 1(d) shows the variation of the voltage across the laser electrodes with increasing OEF strength at different temperatures. This measurement demonstrates the existence of a critical point beyond which the voltage drops down rapidly. At 20 {\textcelsius}, this drop-off point appears at a feedback intensity of 24 dB, while at 30 {\textcelsius}, it appears at 26 dB. Furthermore, the voltage drop rate is faster at lower temperatures. This might be due to the stronger electrical feedback negative signal brought by the high power of the laser at lower temperatures. This drop-off point may be clearly associated with the dynamics transition point of the QD laser. 

\begin{figure*}[t]
\centering{\includegraphics[width=150mm]{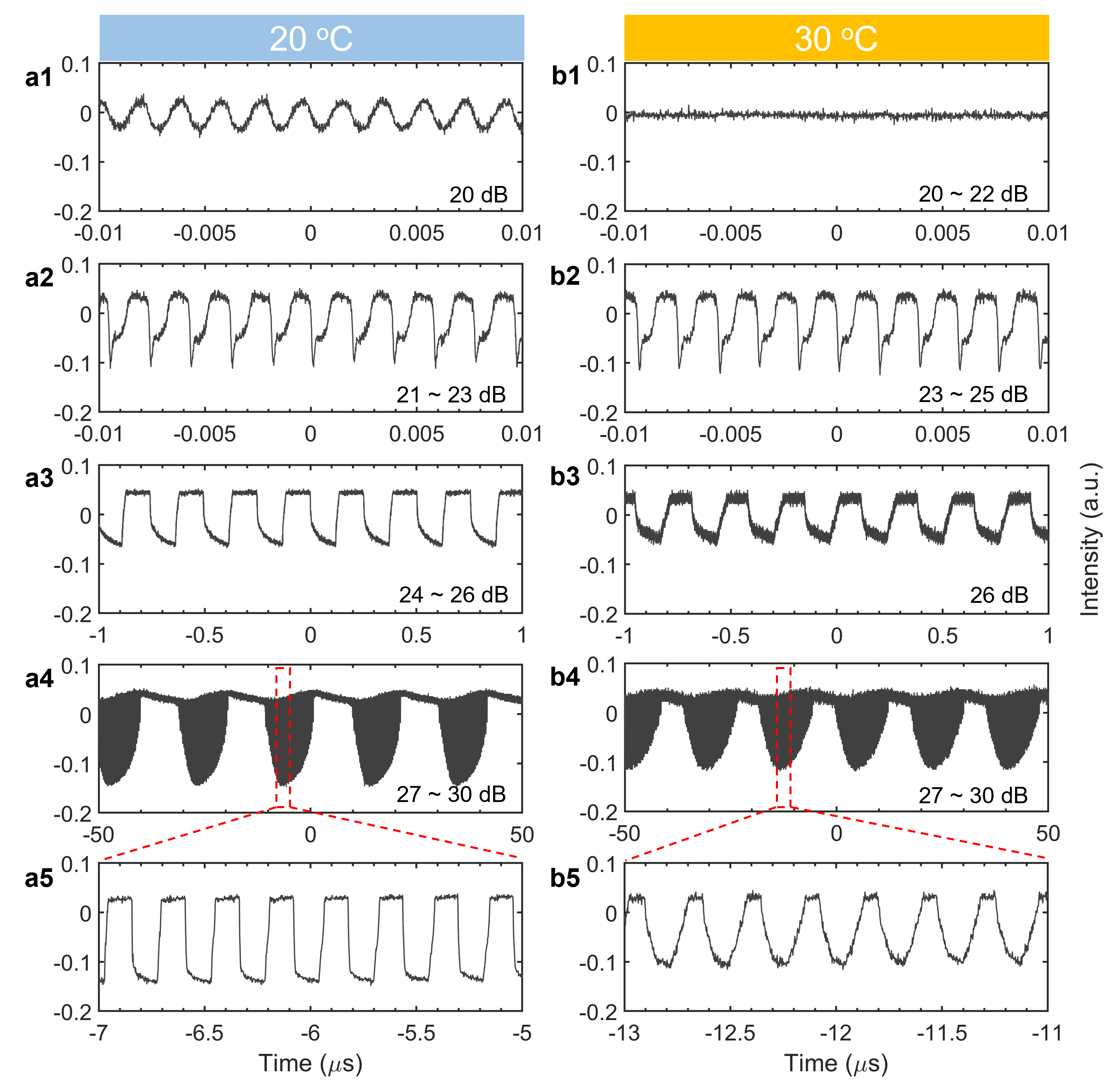}}
\caption{\textbf{Nonlinear dynamics of the QD laser with OEF.} Experimental time dependent output of the QD laser at \textbf{a1}-\textbf{a4} 20 {\textcelsius} and \textbf{b1}-\textbf{b4} 30 {\textcelsius} with different OEF strengths. \textbf{a5} and \textbf{b5} are partial zoom-in views.}
\end{figure*} 

\subsection*{Nonlinear dynamics experiments under OEF} 

Figures 2a and 2b show the electrical spectra of the QD laser under OEF at 20 {\textcelsius} and 30 {\textcelsius}, respectively. At 20 {\textcelsius}, the QD laser begins to show a high-intensity signal at a frequency of 0.5 GHz as the OEF strength increases from 21 dB to 23 dB. When the feedback strength continues to increase, the low-frequency noise below 100 MHz gradually intensifies, as shown in Figures 2a and 2c. Typically, the carriers in semiconductor lasers operating at higher temperatures are subject to thermal effects that alter the carrier transport processes as well as the gain of the laser, which can lead to more complex dynamic changes \cite{patane1999thermal}. As shown in Figure 2b and 2d, the initial oscillation at 0.5 GHz at 20 {\textcelsius} gets much more prominent and even expands on a larger frequency range when the temperature is raised to 30 {\textcelsius}. At the same time, the QD laser exhibits a chaotic-like spectrum state with a bandwidth of about 5 GHz. The low-frequency noise gets stronger at higher feedback strength. Contrasting with conventional optical feedback systems, where QD lasers typically exhibit insensitivity, the experimental procedures employed in this study reveal a distinct characteristic: QD lasers can manifest an abundance of nonlinear dynamics under OEF. Subsequently, we undertake a comprehensive time-domain analysis, by systematically varying feedback strengths and operating temperatures to delve deeper into this intriguing behavior.

\begin{figure*}[t]
\centering{\includegraphics[width=150mm]{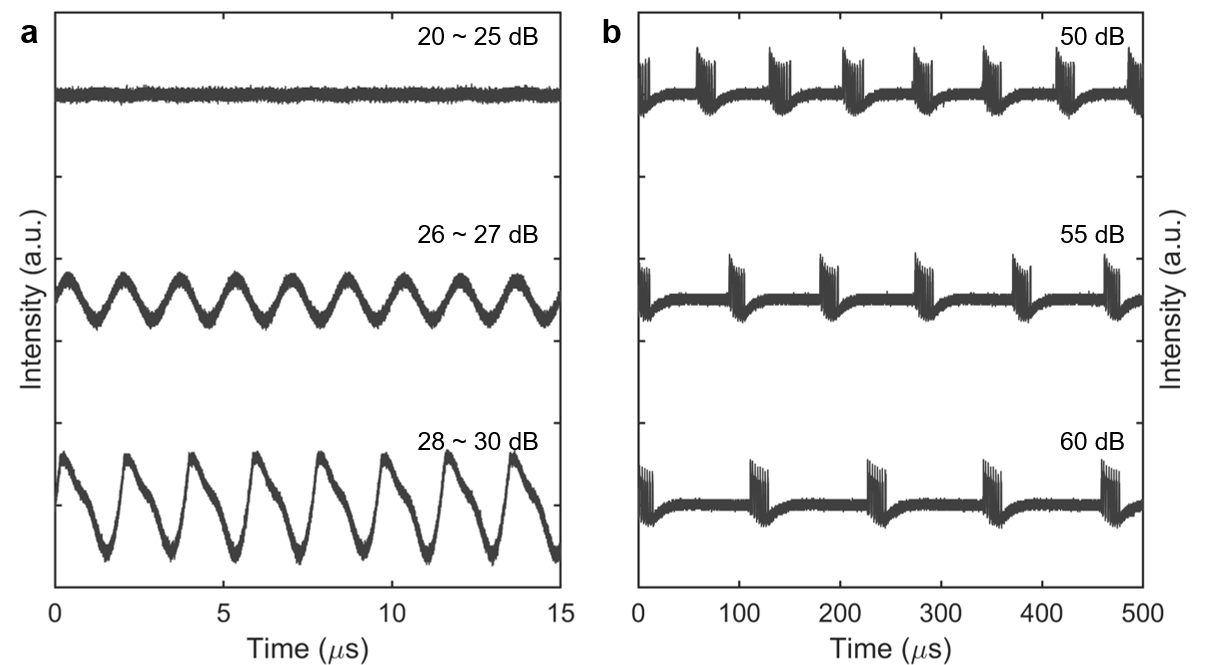}}
\caption{\textbf{Nonlinear dynamics of the QW laser with OEF.} \textbf{a} Experimental time series of QW laser with the same OEF setup as QD laser at different feedback strengths. \textbf{b} Time series of the QW laser after the inclusion of another electrical amplifier with feedback strengths.}
\end{figure*}

Figure 3a1~a5 depict the time-domain signals of the QD laser for different OEF strengths at 20 {\textcelsius}. As the feedback strength increases, the QD laser displays different waveforms ranging from (a1) sine wave, (a2) quasi-square wave, (a3) square wave, to (a4) a mixed state with fast and slow oscillations, where the fast period is shown in (a5). Both the sine wave and the quasi-square wave exhibit a 2 ns period which comes from a high cut-off frequency. Due to the low amplitude of the sine wave, only a weak contribution of the 0.5 GHz peak contribution can be seen in the electrical spectrum, while the quasi-square wave corresponds to stronger frequency peaks and a chaotic-like uplift.  The chaotic-like signal at this point may come from the inconsistent falling edge of the square wave-like. The period of the square wave is 250 ns, which perfectly corresponds to the OEF loop time. As for the mixed state, it is found that the slow period of 20 $\mu$s corresponds to the wave packet occurrence whereas the fast one of 250 ns is linked to the square wave dynamic within the wave packet. As the temperature rises, a greater number of carriers are thermally excited to higher energy states, leading to the possibility of additional alterations in the laser dynamics. Therefore, by increasing the temperature to 30 {\textcelsius}, we found that the sine wave does not appear at lower feedback strengths, but remains steady instead while the appearance of the quasi-square wave occurs much later. Under this temperature, the mixed regime with fast and slow oscillations also appears after the square wave. These regimes displayed in Figure 3b1-b5 are due to the increase of thermal effects which slow down the involvement of carriers in the electrical non-linear changes. 

The QD laser under study outputs a pronounced sensitivity to OEF characterized by intense and sustained dynamic features. As previously mentioned, this rich behavior is fundamentally different from the typical observations encountered in conventional optical feedback systems \cite{duan20191}. To ensure comprehensive analysis, we further investigate the key distinctions by comparing our findings with those obtained from a 1310 nm QW Fabry-Perot laser. By employing an identical setup at a temperature of 30 {\textcelsius}, we generate time series data, which is presented in Figure 4a. It can be found that the QW laser does not exhibit a square wave nor a mixed state with fast and slow oscillations observed even at the highest feedback strength. By varying the OEF strength, different regimes are observed from steady-state to sine wave down to quasi-triangular wave. In order to verify whether the QW laser could produce more dynamics under high feedback strengths, an additional electronic amplifier (RF Bay, LNA-1800, 1 kHz-1.8 GHz, 30 dB gain) was included in the loop. Due to limitations in the availability of electronic attenuators with a broader range, our adjustment options were restricted to a 10 dB attenuation following a 60 dB-RF signal enhancement achieved through the use of two electrical amplifiers, as depicted in Figure 4b. The QW laser exhibits a mixed waveform of fast and slow oscillations under high feedback strengths. As the feedback strength increases, the slow period gradually changes. The fast period, however, remains a square wave with a constant period. This is consistent with the state of the QD laser under high feedback strength with one electronic amplifier. As compared to the QD laser, the QW laser demonstrates extreme OEF insensitivity.

\subsection*{Theoretical modeling and analysis}

\begin{figure*}[t]
\centering{\includegraphics[width=150mm]{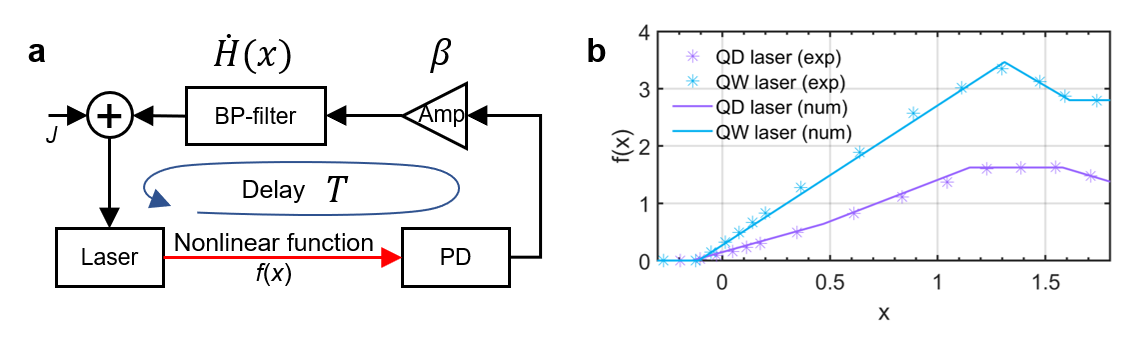}}
\caption{\textbf{Theoretical Modeling of OEF loop.} \textbf{a} Block diagram depicting the time-domain representation of the Ikeda-like OEF loop. The variable $x(t)$ undergoes counterclockwise circulation and interacts with the four key components within the loop. \textbf{b} The experimental and numerical nonlinearity transfer functions of QD (purple) and QW (blue) lasers, respectively.}
\end{figure*}

Our experimental setup can be simplified, as illustrated in Figure 5a, based on the Ikeda-like theory \cite{ikeda1979multiple,chembo2019optoelectronic}. This OEF loop's dynamical properties are governed by the overall bandpass filtering effect hence resulting from the combined bandwidths of the RF amplifier, the PD, the attenuator, and the BT. Exploiting the significant spectral separation between the low cutoff frequency, $f_L$ = 50 kHz, and the high cutoff frequency, $f_H$ = 0.5 GHz, we can model this bandpass filter as a cascade of two first-order linear filters, comprising a high-pass and a low-pass filter. In addition, the loop delay time $T$ = 250 ns determines the period of the time domain signal. The input voltage $V_{\mathrm{in}}(t)$ and output voltage $V_{\mathrm{out}}(t)$ of a cascaded bandpass filter are linked by the equation:

\begin{align}
    \hat{H}(V_{\mathrm{out}}(t))\equiv &\left[ 1+\frac{f_L}{f_H}\right]V_{\mathrm{out}}(t) + \frac{1}{2\pi f_H} \frac{\mathrm{d} V_{\mathrm{out}}(t)}{\mathrm{d} t} \\ \nonumber
    &+ 2\pi f_L\int_{t_0}^{t}V_{\mathrm{out}}(s)ds=V_{\mathrm{in}}(t)
\end{align}

The optical power $P$ at the output of the optical fiber undergoes conversion into an electrical signal through the PD, which follows a time-delayed relationship expressed as $S\times P(t-T)$. Here, $S$ represents the power-dependent responsivity of the PD, and $T$ signifies the time delay resulting from fiber propagation. Importantly, the output voltage of the PD corresponds to the voltage $V_{\mathrm{in}}(t)$ applied to the input of the bandpass filter. Furthermore, the relationship between $V_{\mathrm{RF}}(t)$ and the output voltage can be described as $V_{\mathrm{RF}}(t)=\eta V_{\mathrm{out}}(t)$, where $\eta$ serves as a conversion factor encompassing all linear gains and losses (both electrical and optical) within the feedback loop. Consequently, it can be demonstrated that the radio-frequency voltage $V_{\mathrm{RF}}(t)$ adheres to the following equation:

\begin{align}
    &\left[ 1+\frac{f_L}{f_H}\right]V_{\mathrm{RF}}(t)+\frac{1}{2\pi f_H} \frac{\mathrm{d} V_{\mathrm{RF}}(t)}{\mathrm{d} t}+ 2\pi f_L\int_{t_0}^{t}V_{\mathrm{RF}}(s)ds \\ \nonumber
    &= \eta f_{NL}\left[V_{\mathrm{RF}}(t-T)\right]
\end{align}

The nonlinearity transfer function of the feedback loop denoted as $f_{NL}\left[V_{\mathrm{RF}}(t-T)\right]=S\times P(t-T)$, which is directly dependent on the RF voltage. By taking into account $f_L\ll f_H$, we can reformulate this transfer function in a dimensionless form as follows:

\begin{align}
   x(t) + \tau  \frac{\mathrm{d} x(t)}{\mathrm{d} t} + \frac{1}{\theta} \int^{t}_{t_0} x(s)ds = \eta f_{NL}[x(t-T))]
\end{align}

In this context, we define the system variable as $x(t)=V_{\mathrm{RF}} (t)/V_{\mathrm{REF}}$, where $V_{\mathrm{REF}}$ = 1 V serves as a convenient reference voltage, ensuring that $\left|x(t)\right|=\left|V_{\mathrm{RF}}\right|$. The dimensionless feedback loop linear gain $\eta$. Moreover, we introduce several time parameters: $\tau$ = 1/2$\pi f_H$, and $\theta$ = 1/2$\pi f_L$, which correspond to the high-pass and low-pass cut-off frequencies, respectively.

In most architectures of OEF, the conversion between the electrical and optical signals is typically achieved using a phase or intensity modulator featuring a sinusoidal transfer function. However, a straightforward alternative involves utilizing the seeding laser itself as an electrical-to-optical converter \cite{chengui2020nonlinear}. Building upon this concept, we propose a modeling approach that leverages the inherent effective "elbow" nonlinearity present in the power-intensity transfer function of various types of lasers. This nonlinearity arises due to the laser threshold, carrier leakage, and also the gain compression effects, resulting in a piecewise characteristic. Similarly, the PD can also exhibit a nonlinearity induced by saturation. Therefore, the overall transfer function encompasses a cascade of the nonlinearities of both the laser and PD. In this study, our primary objective is to demonstrate that this cascading effect enables the OEF in QD lasers to attain fast multi-dynamics through a series of bifurcations as the gain is continuously increased. In contrast, QW lasers only exhibit slow oscillations with limited dynamics.

\begin{figure*}[t]
\centering{\includegraphics[width=150mm]{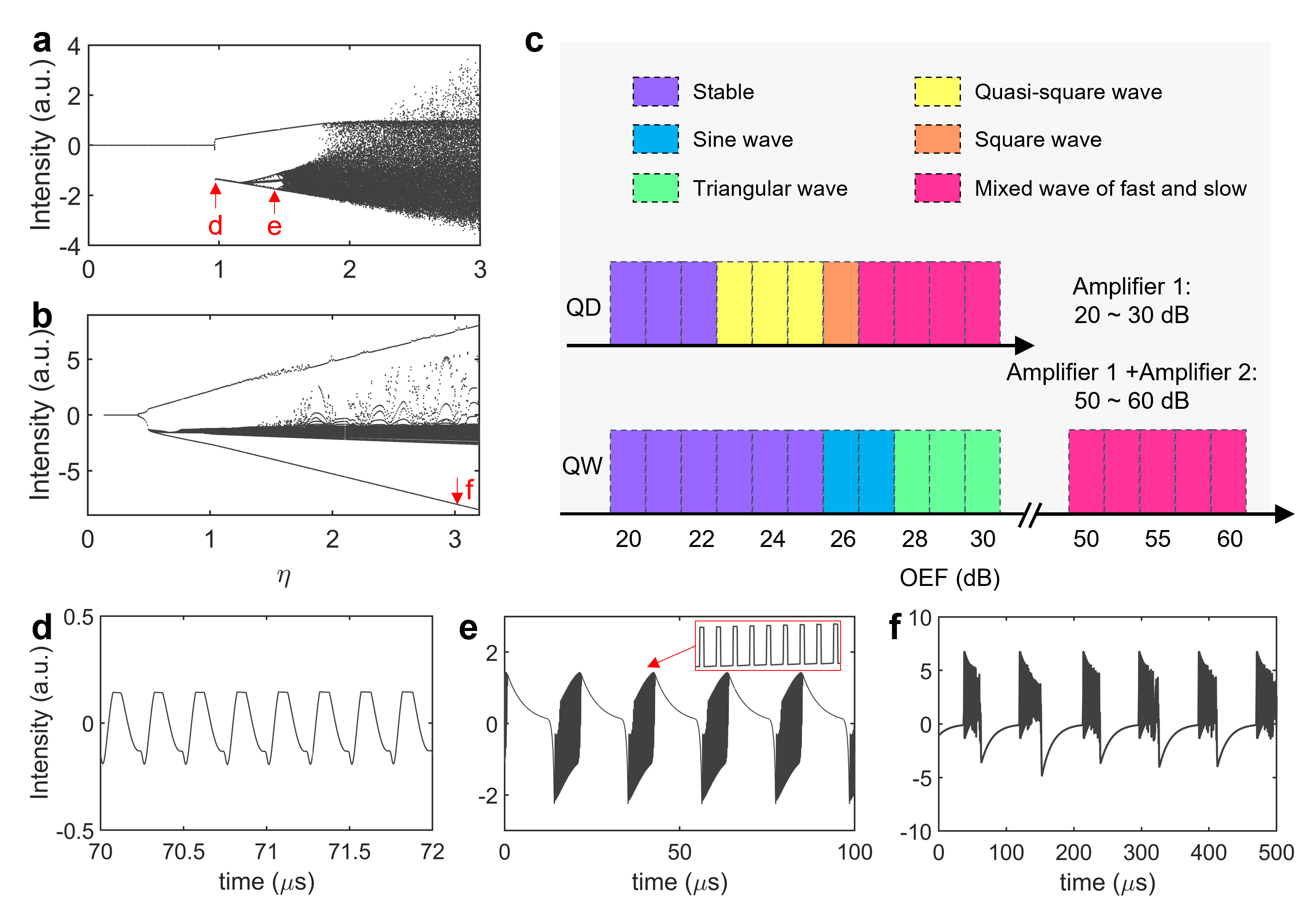}}
\caption{\textbf{Dynamics comparison of QD and QW lasers.} Simulated bifurcation diagrams of \textbf{a} QD and \textbf{b} QW lasers. \textbf{c} Experimental dynamical regions observed in QD and QW lasers, respectively. \textbf{d}-\textbf{f} Simulated time series of the QD laser for a given feedback strength as indicated by the red arrows in \textbf{a}.}
\end{figure*}

To determine the nonlinearity transfer function of QD lasers and QW lasers individually, we conduct experiments using a setup comprising different laser diodes and the same PD in the open-loop configuration. In this setup, an input voltage $V$ is applied to the laser diode, gradually increased, and both this voltage and the corresponding output voltage $V_\mathrm{{PD}}$ at the PD are recorded. These recorded data points are then utilized to generate the plotted points shown in Figure 5b. In fact, the PD remains in the linear respondence region until the lasers reach saturation power (the response curve of the PD can be found on the official website). Therefore, the response curve of the PD in the nonlinearity transfer function can be taken as a linear line. The characteristic inflexion points of the nonlinearity transfer function are mainly determined by the difference in the P-V curves of the lasers. Subsequently, we interpolate these points using piecewise transfer functions for QD laser ($f_{QD}$) and for the QW laser ($f_{QW}$) which read as follows:

\begin{align}
f_{QD}(x) = \begin{cases}
0,\quad &x\leq -0.14 \\
1.05x + 0.15,\quad &-0.14 < x \leq 0.47 \\
1.45x-0.04,\quad &0.47 < x \leq 1.15 \\
1.62,\quad &1.15 < x \leq 1.58 \\
-1.12x + 3.39,\quad &x >1.58 
\end{cases} 
\end{align}
\begin{align}
f_{QW}(x) = \begin{cases}
0,\quad &x\leq -0.11 \\
2.44x+0.27,\quad &-0.11 < x \leq 1.31 \\
-2.2x+6.35,\quad &1.31 < x \leq 1.61 \\
2.80,\quad &x >1.61
\end{cases} 
\end{align}

The coefficients of the piecewise transfer functions are calculated through linear interpolation based on the available experimental data points. The obtained results demonstrate a remarkable level of agreement between the experimental observations and the numerical modeling. The specific segmentation points used in the interpolation process are determined by considering key factors such as the laser threshold, saturation transitions, and saturation drop point. Furthermore, it is important to reiterate that the variable $x$ should not be confused with the input voltage ($V$) applied to the laser diode as recorded in previous measurements. Rather, $x$ refers to the radio-frequency voltage $V_{\mathrm{RF}}$, which is derived from the difference between the input voltage ($V$) and the polarization voltage ($V_{\mathrm{bias}}$) of the laser when operating at twice the threshold current during OEF operation. In the case of the QD laser, the value of $V_{\mathrm{bias}}$ is 1.30 V, while for the QW laser, it is 1.19 V. Moreover, the function $f(x)$ corresponds to the detection voltage ($V_{\mathrm{PD}}$) of the PD. Once all the simulation parameters are obtained, equation (3) can be reformulated as an integrodifferential delay equation (iDDE) \cite{chembo2019optoelectronic,chengui2020nonlinear} such as: 

\begin{align}
\Dot{y} &= x\\
\tau\Dot{x} = -x - &\frac{1}{\theta}y + \eta f_{NL}[x_T]
\end{align}
where $y = \int^{t}_{t_0} x(s)ds$, $x_T \equiv x(t-T)$. $f_{NL}[x_T]$ is represented in detail by equations (4) and (5).

\begin{table*}[htbp]
\caption{Comparison of recent optoelectronic feedback experiments.}
    \label{tab:my_label}
    \centering
    \begin{threeparttable}
    \scalebox{0.8}{
    \begin{tabular}{c|c|c|c|c|c|c}
        \hline 
\multirow{2}*{\makecell[c]{Optical\\ source}} & \multirow{2}*{\makecell[c]{Nonlinear\\ component}\tnote{*}} & \multicolumn{4}{c|}{Generated dynamics} & \multirow{2}*{Ref.}\\
 &  & \makecell[c]{Frequency of\\ sine wave (GHz)} & \makecell[c]{Frequency of\\ square wave (GHz)} & \makecell[c]{Fast-slow\\ oscillation} & \makecell[c]{Chaotic-like\\ state} & \\
\hline 
QW laser & MZM  & n/a\tnote{†}  & n/a  & $\surd$\tnote{‡} & $\surd$ & \cite{kouomou2005chaotic} \\
QW laser & MZM  & n/a  & $2.5\times 10^{-5}$  & $\surd$ & $\surd$ & \cite{mbe2015mixed} \\
QW laser & PM  & n/a  & 0.9  & n/a & $\surd$ & \cite{lavrov2009electro} \\
QW laser & QW laser  & 0.6  & n/a & n/a & $\surd$ & \cite{tang2001chaotic} \\
QW laser & QW laser  & n/a  & 0.12 & n/a & n/a & \cite{islam2021optical} \\
QW laser & QW laser  & $2\times 10^{-3}$  & n/a & $\surd$ & $\surd$ & \cite{chengui2020nonlinear} \\
QD laser & QD laser  & $1.5\times 10^{-3}$  & n/a & n/a & n/a & \cite{munnelly2017chip} \\
QD laser & QD laser  & 0.5  & 0.5  & $\surd$ & $\surd$ & This work \\
\hline 
    \end{tabular}}
\begin{tablenotes}
        \footnotesize
        \item[*] An optoelectronic device with direct feedback of the electrical signal in the loop. 
        \item[†] It means that this dynamic was not presented.
        \item[‡] It means that this dynamic was presented
      \end{tablenotes}
    \end{threeparttable}
\end{table*}

We conduct an investigation and comparison of the dynamic behavior of QD and QW lasers by varying the linear gain ($\eta$), which serves also as the feedback strength, in the feedback loop. The bifurcation diagrams of the QD and QW lasers, obtained by manipulating the feedback strength, are shown in Figures 6a and 6b, respectively. The bifurcation diagram of the QD laser reveals three distinct operational regimes: steady state, single-period region, and multi-period region. Interestingly, the QD laser exhibits a lower feedback strength requirement to enter the multi-period region, characterized by a mixed waveform of fast and slow oscillations, as observed in the experimental results. The simulated time series corresponding to the red arrows in Figures 6d and 6e depict the square wave and mixed waveform of fast and slow oscillations of the QD laser. It is worth noting that the periods and waveforms of these simulated time series align closely with the experimental curves displayed in Figure 3. From the cross-section time domain in Figure 6b, it is evident that the feedback strength goes up to 2.0 where clear fast and slow oscillations appear. Figure 6f shows the time series plots at a feedback strength of 3.1, which perfectly matches the experimental data at high feedback strength in Figure 4. Figure 6c illustrates the experimentally obtained dynamic intervals for the QD and QW lasers at a temperature of 30{\textcelsius}. In particular, the QD laser demonstrates a wide range of dynamics with a single electrical amplifier. On the contrary, the QW laser exhibits stable performance and necessitates higher feedback strength to induce dynamic behavior, corroborating the trends observed in the simulated bifurcation diagram. Additionally, for QD lasers, a fast sine wave and quasi-sine wave emerge within a narrow window between the steady state and the square wave. This observation is primarily evident in experiments conducted at elevated temperatures, possibly attributed to the power decrease at higher temperatures and the resulting reduction in feedback strength variation, which facilitates the adjustment of the feedback strength to this particular point.

\section*{Discussion}
Our results effectively showcase the interplay between the nonlinearity transfer function of semiconductor lasers and the time delay in the context of AC-coupled optoelectronic feedback, giving rise to diverse and complex dynamical regimes. Notably, the PIV characteristics of lasers play a vital role in shaping the nonlinear behavior of the entire OEF loop as highlighted in the existing literature. For instance, previous studies have shown that a simple piecewise transfer function, characterized by zero output below a certain threshold current and linearly increasing above it, can only generate quasi-sinusoidal time traces \cite{chengui2016simplest}. On the other hand, employing a Van der Pol-like nonlinearity transfer function leads to period-doubling routes to chaos \cite{chengui2017dynamics}, indicating that introducing additional nonlinearity can yield novel and complex behaviors not observed in conventional OEF loop. Table 1 compares recent optoelectronic feedback experiments with our work. Recent research has also demonstrated that a nonlinearity transfer function similar to that of QW lasers in our study can only produce slow-fast periodic oscillations at very high feedback strengths (40 V) \cite{chengui2020nonlinear}. Moreover, in alternative OEF loop utilizing a quantum dash DFB laser \cite{islam2020staircase} and a DFB multi-quantum well (MQW) laser \cite{islam2021optical}, only limit cycle oscillations or square wave dynamics. It should be noted that there is just a few MHz high cut-off frequency in the above references. In contrast, when utilizing the QD laser as an optical source, the OEF loop exhibited intense oscillations at a 0.5 GHz cut-off frequency. This frequency significantly surpasses those documented in previous research, demonstrating the exceptional performance of the QD laser source. The absence of a modulation interface in the design of the QD laser could potentially constrain the upper limit of the cut-off frequency. The current implementation of the off-chip OEF loop contributes to extensive latency. The development of future silicon-based on-chip OEF loops could potentially enable the production of square-wave signals in the tens of GHz range. 

Unlike the aforementioned scenarios, QD lasers possess a distinct and more intricate PIV characteristic with multiple piecewise function segments and are more susceptible to saturation compared to QW lasers. This can be attributed to the higher gain compression coefficient in QD lasers (approximately $10^{-16}$ cm$^3$), which is more than one order of magnitude greater than the typical values found in QW lasers (approximately $10^{-17}$ cm$^3$) \cite{bimberg1997ingaas}. The increased gain compression can be associated with the slower relaxation rate of carriers from the wetting layer into the quantum dots \cite{nielsen2004many,ludge2009quantum} owing to the zero-dimensional structure of QDs with discrete energy states. This unique feature significantly limits the number of carriers that can be accommodated in the QDs and restricts the saturation photon number within the cavity, thus suggesting a possible physical explanation for the observed phenomenon. It is noteworthy that the intricate energy level structure of QD lasers compared to QW lasers produces more sophisticated undulating in PIV curves. For instance, if the excited state (ES) of the QD laser participates in emission, thereby inducing the quenching between the ground state (GS) and ES, it results in a PIV curve characterized by numerous threshold points and several saturation zones \cite{roehm2014understanding, huang2018analysis}. In future works, OEF experiments could be conducted utilizing QD lasers that exhibit ES emission. Consequently, this complexity significantly enhances the intricacy of the ensuing nonlinearity transfer function.

In this work, we investigate both experimentally and numerically the response of a silicon-based QD laser under optoelectronic feedback. As OEF strength increases, experiments reveal that the QD laser undergoes various dynamical regimes including, steady state, square waves, and a mixed waveform in which fast and slow oscillations. When the temperature is reduced, we find out that the square wave regime is broadened, which is beneficial for optical switching applications. As compared to QW lasers, QD ones exhibit a larger sensitivity to OEF, which is more conducive to dynamic generation. We confirmed this phenomenon through the integral-differential delay model, obtaining temporal signals consistent with the experimental results. Those dynamics produced by silicon-based QD lasers can be applied to fields, such as optical clocks, optical logic, and optical sensing.
	
\section*{Methods}
\subsection*{Silicon-based QD laser}
The epitaxial layer structure of the QD laser under study is presented in Figure 1(a). The InAs/GaAs Fabry-Perot (FP) laser is 1.35 mm long and has a 3.5 $\mu$m wide ridge waveguide. The active region contains 5 InAs dot layers and p-doped barriers directly grown on an on-axis (001) Si wafer\cite{duan2022four}. The cavity facets have power reflectivity of 60\% and 99\% on the front and rear facets, respectively.

\subsection*{Optoelectronic feedback experiment}
The output light from the QD laser passes through a semiconductor optical amplifier (SOA) set to 150 mA before entering the PD, bringing 10.2 dB of optical amplification. The PD output (Discovery Semiconductor DSC-R401HG, DC–20 GHz) undergoes amplification via a 30 dB-gain amplifier (Photline, 50 kHz–20 GHz) which is followed by a 10 dB attenuator (Telonic Berkeley, DC-2.5 GHz, 1 dB step) prior to reaching the QD laser injection terminals through a Bias Tee (BT, Picosecond 10 kHz–12 GHz bandwidth). There is a power splitter before the bias tee that allows the electrical signal of the loop to be imported to the oscilloscope (20-GHz bandwidth, 40 GS/s).

\subsection*{Numerical modeling of optoelectronic feedback}
The OEF loop can be divided into four basic building blocks according to the Ikeda-like model: a linear gain $\eta$, a nonlinearity transfer function $f_{NL}[x_T]$, a linear integrodifferential operator $\hat{H}(x)$, and a time delay $T$. The system's dynamical properties are governed by the overall bandpass filtering effect hence resulting from the combined bandwidths of the RF amplifier, the PD, the attenuator, and the BT. Exploiting the significant spectral separation between the low cutoff frequency, $f_L$, and the high cutoff frequency, $f_H$, we can model this bandpass filter as a cascade of two first-order linear filters, comprising a high-pass and a low-pass filter.

\section*{Acknowledgments}
The authors acknowledge the financial support of the Institut Mines-Télécom. Shihao Ding’s work is also supported by the China Scholarship Council. Special thanks to Prof. Jia-Ming Liu from the University of California Los Angeles for fruitful discussions. 

\section*{Author contributions}
S. D. and S. Z. conceived the idea. S. D. did experiments and S. Z. did simulations. J. N. provided QD lasers. J. B., H. H., and F. G. performed the manuscript optimization. F. G. supervised the research.

\section*{Conflict of interest} 
The authors declare no competing interests.

\section*{Data Availability} 
The data underlying the results presented in this paper are not publicly available at this time but may be obtained from the authors upon reasonable request.

\bibliography{refs}

\end{multicols}	

\end{document}